\documentclass[%
 reprint,
 amsmath,amssymb,
 aps,
]{revtex4}

\usepackage{color}
\usepackage{graphicx}% Include figure files
\usepackage{dcolumn}% Align table columns on decimal point
\usepackage{bm}% bold math
\begin{document}

\preprint{APS/123-QED}

\title{Atomic indirect measurement and robust binary quantum communication under phase-diffusion noise}% Force line breaks with \\

\author{Min Namkung}
\author{Jeong San Kim}%
 \email{freddie1@khu.ac.kr}
\affiliation{%
Department of Applied Mathematics and Institute of Natural Sciences,
Kyung Hee University, Yongin 17104, Republic of Korea
}%

\date{\today}% It is always \today, today,

\begin{abstract}
It was known that a novel quantum communication protocol surpassing the shot noise 
limit can be proposed by an atomic indirect measurement based on the Jaynes-Cummings model.
Moreover, the quantum communication with the atomic indirect measurement can nearly achieve
the Helstrom bound as well as the accessible information when message is transmitted by an ideal
coherent state. Here, we show that the atomic indirect measurement is robust against the phase-diffusion noise.
By considering the error probability of discriminating received signal, we show that the atomic indirect measurement can also nearly achieve the Helstrom bound as well as the accessible information even the channel is exposed to the phase-diffused noise. Moreover, we further show that atomic indirect measurement outperforms the feedback-excluded receiver composed of a photon number resolving detector and maximum-a-posteriori decision rule when the standard deviation of the phase-diffusion channel is not too large. 
\end{abstract}

\maketitle

%\tableofcontents

\section{Introduction}
Over several decades, communication protocols with conventional measurements such as a homodyne measurement \cite{m.ban0}  and a direct measurement  \cite{k.tsujino} have been widely used for sharing information between distant parties \cite{j.g.proakis,i.a.burenkov}. Although the conventional measurements can be easily understood and implented simply by means of the electromagnetic
theory \cite{m.fox}, conventional measurements can never surpass the shot noise limit (standard quantum limit) due
to the statistical and physical properties of the optical signal and the measurement device \cite{i.a.burenkov}.

In quantum theory, the shot noise limit can be surpassed by novel and unconventional measurements using a
displacement operation, a photon detector, and real-time electric feedback \cite{g.cariolaro,r.s.kennedy}. It is also shown that
a certain type of unconventional measurement with real-time electric feedback \cite{s.j.dolinar} can reach the Helstrom bound when the message is encoded in one of two ideal coherent states \cite{r.j.glauber}. Unfortunately, the real-time electric feedback is generally hard to be implemented in practical quantum communication \cite{g.cariolaro}. 

For this reason, several unconventional measurements have been proposed without real-time electric feedback \cite{m.sasaki,k.tsujino,s.izumi,r.han,m.namkung,m.namkung2,m.t.dimario}. Recently, it has been shown that the atomic indirect measurement \cite{r.han,m.namkung,m.namkung2} can nearly achieve the Helstrom bound when a message is encoded in one of several standard \cite{r.j.glauber} or non-standard coherent states \cite{e.m.f.curado}. However, these works only consider the case of ideal quantum channels between a sender and a receiver, which is hard to be assumed in practice. Thus, it is natural to consider and analyze the quantum communication protocol performed under the existence of noise. Recently, phase-diffusion noise has been considered in several quantum communication protocols \cite{s.olivares,m.t.dimario2}. 

Here, we show that the atomic indirect measurement is robust against the phase-diffusion noise.
By considering the error probability of discriminating received signal, we show that the atomic indirect measurement can also nearly achieve the Helstrom bound as well as the accessible information even the channel is exposed to the phase-diffused noise. Moreover, we further show that atomic indirect measurement outperforms the feedback-excluded receiver \cite{m.t.dimario,m.t.dimario2} composed of a photon number resolving detector and maximum-a-posteriori decision rule when the standard deviation of the phase-diffusion channel is not too large. As the situation here deals with noisy quantum communication channels, our result is more relevant to the general quantum communication protocols than the previous results \cite{r.han,m.namkung,m.namkung2}. 

The article is organized as follows. In Section II, we briefly introduce the preliminaries regarding the binary quantum communication scheme. In Section III, we consider the minimum value of the error probability and the maximum value of the Shannon mutual information as figures of merit for a quantum communication protocol under phase-diffused noise. In Section IV, we show that the atomic indirect measurement is still robust against the phase-diffusion noise in terms of the figures of merit obtained in Secton III. Moreover, we also show that the atomic indirect measurement outperforms the feedback-excluded receiver \cite{m.t.dimario,m.t.dimario2}. Finally, we conclude our results in Section V.

\section{Quantum communication protocol under phase-diffusion noise}

\subsection{Binary quantum communication protocol}
Suppose that Alice (sender) prepares one bit message $x\in\{1,2\}$ with a prior probability $q_x\in\{q_1,q_2\}$ and encodes it in a quantum state $\rho_x\in\{\rho_1,\rho_2\}$. In practical quantum communication, a quantum channel $\mathcal{N}$ transforms the initial quantum state $\rho_x$ to a noisy quantum state $\mathcal{N}(\rho_x)$, then Bob (receiver) performs a quantum measurement on the quantum state $\mathcal{N}(\rho_x)$ to recover the message $x$.  If Bob's measurement is described by a positive-operator-valued-measurement (POVM) $\mathcal{M}=\{M_1,M_2\}$, the conditional probability that Bob obtains a measurement outcome $y\in\{1,2\}$ is described by
\begin{equation}
\mathrm{Pr}(y|x)=\mathrm{Tr}\left\{\mathcal{N}(\rho_x)M_y\right\}.\label{born}
\end{equation} 
Using Eq. (\ref{born}), \textit{the error probability} that Bob's measurement incorrectly decides the Alice's message can be given by
\begin{equation}
P_e(\mathcal{E},\mathcal{M})=\sum_{x\not=y}q_x\mathrm{Tr}\left\{\mathcal{N}(\rho_x)M_y\right\}=1-\sum_{x=1}^{2}\mathrm{Tr}\left\{\mathcal{N}(\rho_x)M_x\right\},\label{p_e}
\end{equation}
where $\mathcal{E}=\{q_x,\mathcal{N}(\rho_x)\}_{x=1}^{2}$ is the ensemble composed of the transformed quantum states $\mathcal{N}(\rho_x)$ with the corresponding prior probability $q_x$. 

The minimum value of $P_e(\mathcal{E},\mathcal{M})$ in Eq. (\ref{p_e}) over all possible POVM $\mathcal{M}$ of Bob:
\begin{equation}
P_{hel}(\mathcal{E})=\min_{\mathcal{M}}P_e(\mathcal{E},\mathcal{M}),\label{hel}
\end{equation}
is called \textit{the Helstrom bound} \cite{c.w.helstrom}. Moreover, the amount of information shared between Alice and Bob is given by   \cite{t.cover}
\begin{equation}
I(\mathcal{E},\mathcal{M})=\sum_{x,y=1}^{2}\mathrm{Pr}(x,y)\log_2\frac{\mathrm{Pr}(x,y)}{q_x\mathrm{Pr}(y)},\label{mi}
\end{equation}
where 
\begin{equation}
\mathrm{Pr}(x,y)=q_x\mathrm{Tr}\{\mathcal{N}(\rho_x)M_y\},\label{pr_xy}
\end{equation}
is a joint probability that Alice prepares a binary message $x\in\{1,2\}$ and Bob obtains a measurement outcome $y$ from the POVM $\mathcal{M}=\{M_1,M_2\}$. 
\begin{equation}
\mathrm{Pr}(y)=\sum_{x}\mathrm{Pr}(x,y), \label{pr_y}
\end{equation}
is a marginal probability that Bob obtains a measurement outcome $y$. The quantity in Eq. (\ref{mi}) is also referred to as \textit{the Shannon mutual information}. The maximum value of Eq.(\ref{mi}) over all possible POVM $\mathcal{M}$ of Bob:
\begin{equation}
I_{acc}(\mathcal{E})=\max_{\mathcal{M}}I(\mathcal{E},\mathcal{M}),\label{acc}
\end{equation}
is called \textit{the accessible information}. Thus, the efficiency of a quantum channel can be maximized by constructing POVMs realizing the Helstrom bound or the accessible information.

Due to the optimization problem over all possible POVMs, the analytic evaluation of the accessible information in Eq. (\ref{acc}) is generally hard except for two qubit pure states \cite{r.han2} and some simple case of three pure states \cite{d.kaszlikowski}. Moreover, it is also known that the POVM realizing the Helstrom bound is not always equal to that of the accessible information \cite{m.ban,d.kaszlikowski}.

\subsection{Phase-diffusion noise}
In quantum communication, the \textit{phase-diffusion noise} is described as the following quantum channel \cite{s.olivares}
\begin{equation}
\mathcal{N}_{\sigma}(\cdot)=\frac{1}{\sqrt{2\pi}\sigma}\int_{-\infty}^{+\infty}d\phi e^{-\frac{\phi^2}{2\sigma^2}}U(\phi)(\cdot)U^\dagger(\phi),\label{pd}
\end{equation}
where $\sigma$ is the standard deviation of the phase $\phi\in\mathbb{R}$, $U(\phi)$ is a unitary operator
\begin{equation}
U(\phi):=e^{i\phi a^\dagger a}.\label{u}
\end{equation}
Here, $a$ and $a^\dagger$ are annihilation and creation operators, respectively, such that 
\begin{align}
 a|n\rangle&=\sqrt{n}|n-1\rangle, \ \forall n\in\mathbb{Z}^+,\nonumber\\
 a^\dagger |n\rangle&=\sqrt{n+1}|n+1\rangle, \ \forall n\in\mathbb{Z}^+\cup\{0\},\label{a}
\end{align}
with the Fock basis $\{|n\rangle:n\in\mathbb{Z}^+\cup\{0\}\}$.  

If Alice encodes her message $x\in\{1,2\}$ in an ideal coherent state \cite{r.j.glauber}
\begin{equation}
|\alpha_x\rangle=e^{-\frac{1}{2}|\alpha_x|^2}\sum_{n=0}^{\infty}\frac{\alpha_x^n}{\sqrt{n!}}|n\rangle, \ \ \alpha_x\in\mathbb{R},\label{id_c}
\end{equation}
then the phase-diffusion noise in Eq. (\ref{pd}) transforms the ideal coherent state in Eq. (\ref{id_c}) to
\begin{equation}
\tau_x:=\mathcal{N}_\sigma(|\alpha_x\rangle\langle\alpha_x|)=\frac{1}{\sqrt{2\pi}\sigma}\int_{-\infty}^{+\infty}d\phi e^{-\frac{\phi^2}{2\sigma^2}}|\alpha_x e^{i\phi}\rangle\langle\alpha_x e^{i\phi}|=e^{-\alpha_x^2}\sum_{n,m=0}^{\infty}\frac{e^{-\frac{1}{2}\sigma^2(n-m)^2}}{\sqrt{n!m!}}\alpha_x^{n+m}|n\rangle\langle m|.\label{pd_c}
\end{equation}
The state Eq. (\ref{pd_c}) is called \textit{the phase-diffused coherent state}. The rest of this section is to evaluate the Helstrom bound and the accessible information of the phase-diffusion noise described as a quantum channel in Eq. (\ref{pd}).

\subsubsection{Helstom bound}
Suppose that Alice encodes a binary message $x\in\{1,2\}$ to a coherent state $|\alpha_x\rangle$ with a prior probabilities $q_x$, and sends it to Bob through a phase-diffusion channel $\mathcal{N}_\sigma$. To decode the message transmitted as the phase-diffused coherent state $\tau_x=\mathcal{N}_\sigma(|\alpha_x\rangle\langle\alpha_x|)$ in Eq. (\ref{pd_c}), Bob needs to perform the measurement on $\tau_x$. The Helstrom bound for these binary phase-diffused coherent states $\mathcal{E}_\sigma=\{q_x,\tau_x\}_{x=1}^{2}$ is given by \cite{s.olivares}
\begin{equation}
P_{hel}(\mathcal{E}_\sigma)=\frac{1}{2}-\frac{1}{2}\left\Vert\Lambda_\sigma\right\Vert_1,
\end{equation}
where $\Lambda_\sigma$ is a Hermitian operator
\begin{equation}
\Lambda_\sigma=\sum_{n,m=0}^{\infty}\frac{e^{-\frac{1}{2}\sigma^2(n-m)^2}}{\sqrt{n!m!}}\left(q_1e^{-\alpha_1^2}\alpha_1^n\alpha_1^m-q_2e^{-\alpha_2^2}\alpha_2^n\alpha_2^m\right)|n\rangle\langle m|.
\end{equation}

\subsubsection{Accessible information}
Although the the analytical form of fhe accessible information is not known except for the case that binary pure states are discriminated by 2 element POVM \cite{r.han2}, we can efficiently approximate the accessible information by using the steepest ascent method \cite{j.rehacek}; let us consider the Hermitian operator defined as
\begin{equation}
R_y=\sum_{x=1}^{2}q_x\tau_x\log_2\left(\frac{\mathrm{Pr}(x,y)}{q_x\mathrm{Pr}(y)}\right), \ \ y=1,2
\end{equation}
where $\mathrm{Pr}(x,y)$ is the joint probability in Eq. (\ref{pr_xy}) and $\mathrm{Pr}(y)$ is the marginal probability in Eq. (\ref{pr_y}) when Bob performs the POVM $\mathcal{M}=\{M_1,M_2\}$. Now, we perform the iterations composed of the following two steps, 
\begin{itemize}
\item Define $G_y=\mathbb{I}+\lambda(R_y-\sum_{z=1}^{2}R_zM_z)$ and compute $\widetilde{M}_y=G_y^\dagger M_yG_y$, where $\mathbb{I}$ is an identity operator and $\lambda$ is a small positive number.
\item Update new POVM elements as $S^{-1/2}\widetilde{M}_yS^{-1/2}\rightarrow M_y$, where $S=\sum_{z=1}^{2}\widetilde{M}_{z}$.  
\end{itemize}

It is noted that newly updated POVM converges to a POVM satisfying the stationary conditions \cite{m.ban,j.rehacek}
\begin{equation}
M_y\Gamma=M_yR_y, \ \ R_yM_y=\Gamma M_y,\label{nec}
\end{equation}
where
\begin{equation}
\Gamma=\sum_yR_yM_y=\sum_yM_yR_y.
\end{equation}
Eq. (\ref{nec}) is a necessary condition that the POVM $\mathcal{M}=\{M_1,M_2\}$ maximizes the Shannon mutual information of Eq.(\ref{mi}).

Because $\tau_x$ is an infinite-dimensional quantum state, it also needs to be approximated onto some finite-dimensional quantum system as  \cite{s.guerrini}
\begin{equation}
\tau_x\approx e^{-\alpha_x^2}\sum_{n=0}^{N}\sum_{m=0}^{M}\frac{e^{-\frac{1}{2}\sigma^2(n-m)^2}}{\sqrt{n!m!}}\alpha_x^{n+m}|n\rangle\langle m|,\label{tau_app}
\end{equation}
for sufficiently large $N$ and $M$.

\begin{figure*}
\centering
\includegraphics[scale=0.55]{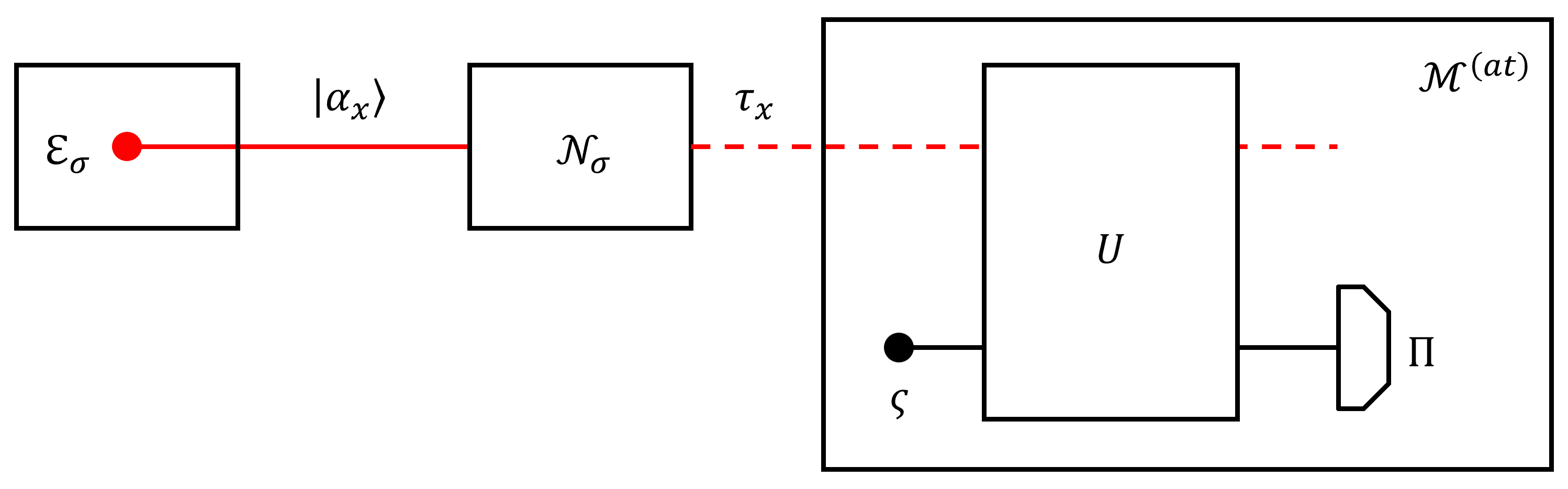}
\caption{Schematic of the quantum communication protocol with an atomic indirect measurement under a phase-diffusion channel $\mathcal{N}$. Here, $\mathcal{E}_\sigma$ is an ensemble of Alice, $\tau_x=\mathcal{N}_\sigma(|\alpha_x\rangle\langle\alpha_x|)$ is a phase-diffused coherent state, and $\mathcal{M}^{(at)}$ is an atomic indirect measurement. In $\mathcal{M}^{(at)}$, $\varsigma=|g\rangle\langle g|$ is an initial (ground) atomic state on a Hilbert space $\widetilde{\mathcal{H}}$, $\Pi$ is an atomic projective measurement, and $U$ is a unitary operator. }
\end{figure*}

\section{Binary quantum Communication protocol \\ with atomic indirect measurement}

The atomic indirect measurement is described by the four elements \cite{m.ozawa}
\begin{equation}
\{\widetilde{\mathcal{H}},\varsigma,\Pi,U\},\label{si}
\end{equation}
where $\widetilde{\mathcal{H}}$ is a Hilbert space describing a two-level atom, $\varsigma$ is the ground state of the two-level atom, $\Pi=\{\Pi_1,\Pi_2\}$ is an projective measurement on the two-level atom, and $U$ is the unitary operator describing the interaction between the light and the two-level atom. The atomic indirect measurement is illustrated in Fig. 1.

By denoting $|g\rangle$ and $|e\rangle$ as the ground and excited states of the two-level atom, respectively, we have 
\begin{equation}
\widetilde{\mathcal{H}}=\mathrm{span}\{|g\rangle,|e\rangle\}, \ \ \varsigma=|g\rangle\langle g|.
\end{equation}
Moreover, the elements of $\Pi$ can also be written as $\Pi_y=|\pi_y\rangle\langle\pi_y|$ where
\begin{equation}
|\pi_1\rangle=\cos\theta|g\rangle+e^{i\xi}\sin\theta|e\rangle, \ \ |\pi_2\rangle=\sin\theta|g\rangle-e^{i\xi}\cos\theta|e\rangle.\label{pi}
\end{equation}

We consider the atomic indirect measurement based on the Jaynes-Cummings model \cite{e.t.jaynes} where the interaction between the light and the two-level atom is described by the unitary operator \cite{m.namkung3}
\begin{equation}
U(t)=\exp\left\{-i\Phi(t)(a\otimes\sigma_++a^\dagger\otimes\sigma_-)\right\},\label{u(t)}
\end{equation}
with non-Hermitian operators
\begin{equation}
\sigma_+=|e\rangle\langle g|, \ \ \sigma_-=|g\rangle\langle e|,
\end{equation}
and the time-dependent real parameter $\Phi=\Phi(t)$. If the atomic projective measurement is performed at time $T$, the unitary operator $U$ in Eq. (\ref{si}) is given by $U=U(T)$.  

After the phase-diffused coherent state $\tau_x$ in Eq. (\ref{pd_c}) interacts with the two-level atom by the unitary operator $U(t)$, the state of the composite system becomes 
\begin{equation}
U(t)\left(\tau_x\otimes\varsigma\right) U^\dagger (t).
\end{equation}
If Bob obtains an outcome $y\in\{1,2\}$ by measuring the two-level atom, then the phase-diffused coherent state is transformed to the (unnormalized) quantum state
\begin{equation}
\mathrm{Tr}_{A}\left\{\left(\mathbb{I}_{L}\otimes\Pi_y\right)U(t)\left(\tau_x\otimes\varsigma\right)U^\dagger(t)\left(\mathbb{I}_{L}\otimes\Pi_y\right)^\dagger\right\}= K_y\tau_xK_y^\dagger,
\end{equation}
where $\mathrm{Tr}_A$ denotes the partial trace over the atom system, $\mathbb{I}_L$ denotes the identity operator of the light system, and $\{K_1,K_2\}$ is the set of the Kraus operators on the light system \cite{m.namkung3}
\begin{align}
&K_1=\cos\theta\cos\{\Phi\sqrt{a^\dagger a}\}-ie^{-i\xi}\sin\theta\sum_{n=0}^{\infty}\frac{(-1)^n}{(2n+1)!}\Phi^{2n+1}a(a^\dagger a)^n,\nonumber\\
&K_2=\sin\theta\cos\{\Phi\sqrt{a^\dagger a}\}+ie^{-i\xi}\cos\theta\sum_{n=0}^{\infty}\frac{(-1)^n}{(2n+1)!}\Phi^{2n+1}a(a^\dagger a)^n.\label{k_at}
\end{align}
In other words, the atomic indirect measurement can be described by a POVM $\mathcal{M}^{(at)}=\{K_1^\dagger K_1,K_2^\dagger K_2\}$ of the light system. These Kraus operators are used for deriving both the error probability and the Shannon mutual information. 

We first evaluate the error probability of discriminating the phase-diffused coherent states given by $\mathcal{E}_{\sigma}=\{q_x,\tau_x\}_{x=1}^{2}$ From the definition of the error probability in Eq. (\ref{p_e}), it is straightforward to show that the error probability for the Kraus operators given by Eq. (\ref{k_at}) is \begin{align}
P_e(\mathcal{E}_\sigma,\mathcal{M}^{(at)})&=1-\frac{q_1}{\sqrt{2\pi}\sigma}\int_{-\infty}^{+\infty}d\phi e^{-\frac{\phi^2}{2\sigma^2}}\left\Vert K_1|\alpha_1e^{i\phi}\rangle\right\Vert^2-\frac{q_2}{\sqrt{2\pi}\sigma}\int_{-\infty}^{+\infty}d\phi e^{-\frac{\phi^2}{2\sigma^2}}\left\Vert K_2|\alpha_2e^{i\phi}\rangle\right\Vert^2,\nonumber\\
&=1-q_1e^{-\alpha_1^2}f_1^{(\sigma)}(\xi,\theta,\Phi)-q_2e^{-\alpha_2^2}f_2^{(\sigma)}(\xi,\theta,\Phi),\label{p_e_pd}
\end{align}
where $f_x^{(\sigma)}(\xi,\theta,\Phi)$ are defined as
\begin{align}
f_1^{(\sigma)}&(\xi,\theta,\Phi)\nonumber\\
&=\sum_{n=0}^{\infty}\left\{\frac{\alpha_1^{2n}}{n!}\cos^2\theta\cos^2(\Phi\sqrt{n})+\frac{\alpha_1^{2(n+1)}}{(n+1)!}\sin^2\theta\sin^2(\Phi\sqrt{n+1})+\frac{\alpha_1^{2n+1}e^{-\frac{1}{2}\sigma^2}}{\sqrt{n!(n+1)!}}\sin\xi\sin2\theta\cos(\Phi\sqrt{n})\sin(\Phi\sqrt{n+1})\right\},\nonumber\\
f_2^{(\sigma)}&(\xi,\theta,\Phi)\nonumber\\
&=\sum_{n=0}^{\infty}\left\{\frac{\alpha_2^{2n}}{n!}\sin^2\theta\cos^2(\Phi\sqrt{n})+\frac{\alpha_2^{2(n+1)}}{(n+1)!}\cos^2\theta\sin^2(\Phi\sqrt{n+1})-\frac{\alpha_2^{2n+1}e^{-\frac{1}{2}\sigma^2}}{\sqrt{n!(n+1)!}}\sin\xi\sin2\theta\cos(\Phi\sqrt{n})\sin(\Phi\sqrt{n+1})\right\}.\label{f1f2}
\end{align}
Here, the following integration is used to derive Eqs. (\ref{p_e_pd}) and (\ref{f1f2}):
\begin{equation}
\int_{-\infty}^{+\infty}d\phi e^{-\frac{\phi^2}{2\sigma^2}}\sin(\phi-\xi)=\sqrt{2\pi}\sigma e^{-\frac{1}{2}\sigma^2}\sin\xi.
\end{equation}
We note that Eq. (\ref{p_e_pd}) encapsulates the result of Ref. \cite{r.han} as a special case that $\sigma$ is equal to zero. 

Now, we evaluate the Shannon mutual information shared between Alice and Bob by discriminating the binary phase-diffused coherent states by the atomic indirect measurement. From the phase-diffused coherent states in Eq. (\ref{pd_c}) and the Kraus operators in Eq. (\ref{k_at}), the joint probability $\mathrm{Pr}(x,y)$ in Eq. (\ref{mi}) is straightforwardly evaluated as
\begin{align}
&\mathrm{Pr}(1,1)=q_1e^{-\alpha_1^2}f_1^{(\sigma)}(\xi,\theta,\Phi), \ \ \mathrm{Pr}(1,2)=q_1e^{-\alpha_1^2}\bar{f}_1^{(\sigma)}(\xi,\theta,\Phi),\nonumber\\
&\mathrm{Pr}(2,1)=q_2e^{-\alpha_2^2}\bar{f}_2^{(\sigma)}(\xi,\theta,\Phi), \ \ \mathrm{Pr}(2,2)=q_2e^{-\alpha_2^2}f_2^{(\sigma)}(\xi,\theta,\Phi),\label{jp}
\end{align}
where $f_x^{(\sigma)}(\xi,\theta,\Phi)$ are defined in Eq. (\ref{f1f2}) and $\bar{f}_x^{(\sigma)}(\xi,\theta,\Phi)$ are defined as
\begin{align}
\bar{f}_1^{(\sigma)}&(\xi,\theta,\Phi)\nonumber\\
&=\sum_{n=0}^{\infty}\left\{\frac{\alpha_1^{2n}}{n!}\sin^2\theta\cos^2(\Phi\sqrt{n})+\frac{\alpha_1^{2(n+1)}}{(n+1)!}\cos^2\theta\sin^2(\Phi\sqrt{n+1})-\frac{\alpha_1^{2n+1}e^{-\frac{1}{2}\sigma^2}}{\sqrt{n!(n+1)!}}\sin\xi\sin2\theta\cos(\Phi\sqrt{n})\sin(\Phi\sqrt{n+1})\right\},\nonumber\\
\bar{f}_2^{(\sigma)}&(\xi,\theta,\Phi)\nonumber\\
&=\sum_{n=0}^{\infty}\left\{\frac{\alpha_2^{2n}}{n!}\cos^2\theta\cos^2(\Phi\sqrt{n})+\frac{\alpha_2^{2(n+1)}}{(n+1)!}\sin^2\theta\sin^2(\Phi\sqrt{n+1})+\frac{\alpha_2^{2n+1}e^{-\frac{1}{2}\sigma^2}}{\sqrt{n!(n+1)!}}\sin\xi\sin2\theta\cos(\Phi\sqrt{n})\sin(\Phi\sqrt{n+1})\right\}.\label{f1f2bar}
\end{align}

As shown in Eqs. (\ref{p_e_pd})-(\ref{f1f2}) and Eqs. (\ref{jp})-(\ref{f1f2bar}), the error probability and the Shannon mutual information are the nonlinear functions of the three parameters $(\xi,\theta,\Phi)$. To use the conventional method of optimization, we need to find the vanishing gradients of the error probability and the Shannon mutual information. As this imposes us to deal with a system of nonlinear equations, it is hard to analytically find the vanishing gradients. In next section, we try the numerical method to optimize the error probability and the Shannon mutual information over all possible atomic indirect measurements

\section{numerical optimization}
We numerically obtain the minimum value of the error probability and the maximum value of the Shannon mutual information over all possible atomic indirect measurements. Here, the average of the mean photon number of the ensemble $\mathcal{E}_\sigma=\left\{q_x,\tau_x\right\}_{x=1}^{2}$ of Alice is given by \cite{m.t.dimario2}
\begin{equation}
\bar{n}=q_1\mathrm{Tr}\left\{\tau_1a^\dagger a\right\}+q_2\mathrm{Tr}\left\{\tau_2a^\dagger a\right\}=q_1\alpha_1^2+q_2\alpha_2^2,
\end{equation}
where $a$ and $a^\dagger$ are annihilation and creation operators, respectively, and $\alpha_x$ is the amplitude of the phase-diffused coherent state $\tau_x$ in Eq. (\ref{pd_c}). 

For a fixed $\bar{n}$, we consider the following two cases: The first case is when $\alpha_1=0$ or $\alpha_2=0$. In this case, $\mathcal{E}_\sigma$ represents on-off keying (OOK) signal. The second case is when $\alpha_1=\alpha_2$. In this case, $\mathcal{E}_\sigma$ represents binary phase shift keying (BPSK) signal. We compare the minimum value of the error probability and the maximum value of the Shannon mutual information over all possible atomic indirect measurements with these of the feedback-excluded receivers composed of a photon number-resolving detector and maximum-a-posteriori decision rule \cite{m.t.dimario,m.t.dimario2}. Throughout this section, we consider the case of $q_1=q_2$, that means equiprobable input signals.

\newpage
\subsection{Numerical minimum value of error probability}
\begin{figure*}
\centering
\includegraphics[scale=0.42]{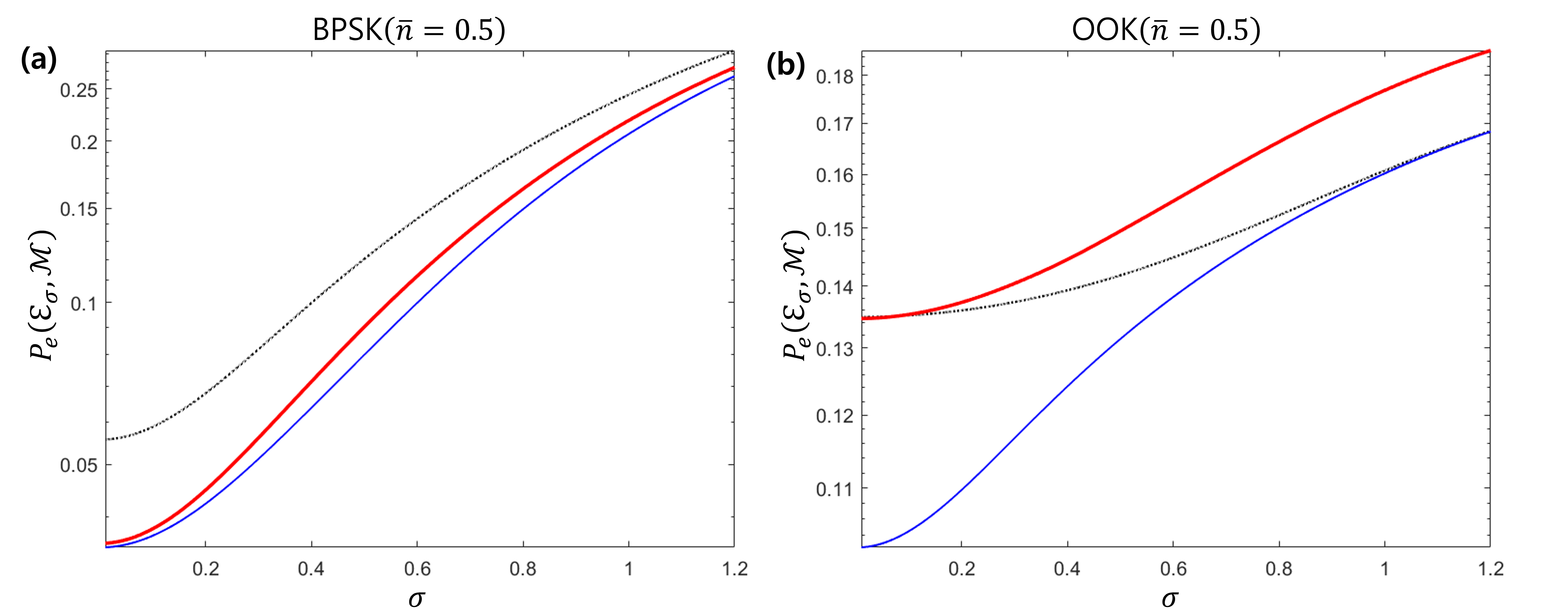}
\caption{The minimum error probability of the atomic indirect measurement when $\bar{n}=0.5$ in the region of $0\le\sigma\le1.2$. In this figure, BPSK signal is considered in Fig. 2(a) whereas OOK signal is in Fig. 2(b). The thick solid red lines show the minimum error probability of the atomic indirect measurement. The dotted black lines show the minimum error probability of the feedback-excluded receiver that the photon number resolution is equal to one and the invisibility is given by 0.998. The solid blue lines show the Helstrom bound.}
\end{figure*}
Fig. 2 depicts the minimum error probability of the atomic indirect measurement when $\bar{n}=0.5$ in the region of $0\le\sigma\le1.2$. In this figure, BPSK signal is considered in Fig. 2(a) whereas OOK signal is in Fig. 2(b). The thick solid red lines show the minimum error probability of the atomic indirect measurement. The dotted black lines show the minimum error probability of the feedback-excluded receiver that the photon number resolution is equal to one and the invisibility is given by 0.998 \cite{m.t.dimario,m.t.dimario2}. The solid blue lines show the Helstrom bound.

According to Fig. 2(a), the minimum error probability of the atomic indirect measurement is closer to the Helstrom bound than that of the feedback-excluded receiver. In Fig. 2(b), the minimum error probability of the feedback-excluded receiver is closer to the Helstrom bound in most range of $\sigma$.  \textit{This result means that the atomic indirect measurement outperforms the feedback-excluded receiver when BPSK signal was sent through the phase-diffused quantum channel instead of OOK signal.} For this reason, we mainly focus on the quantum communication protocol with BPSK signal.

\newpage
\begin{figure*}
\includegraphics[scale=0.42]{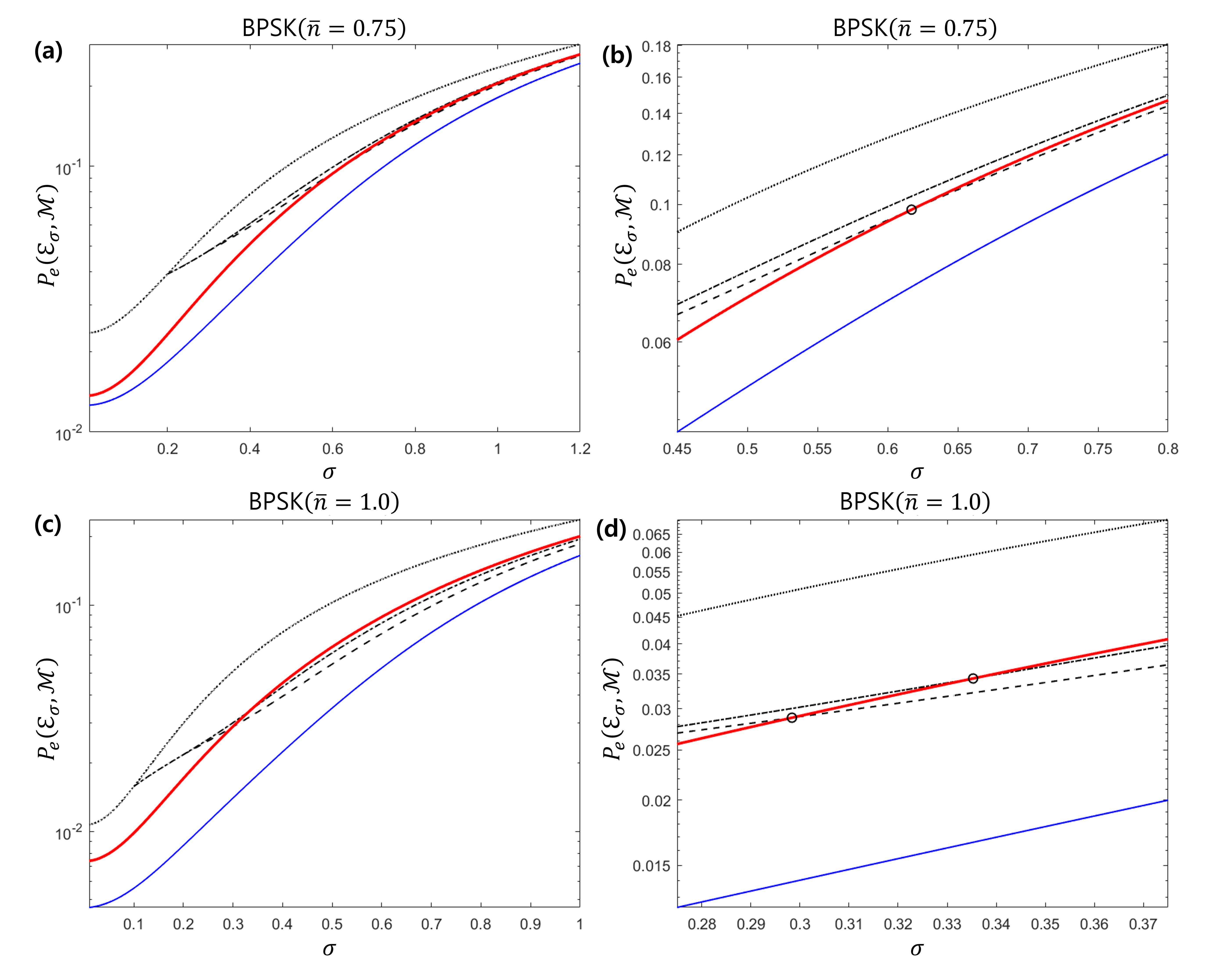}
\caption{The minimum error probability of the atomic indirect measurement when the message is encoded in BPSK signal. Here, we consider the case that $\bar{n}=0.75$ in the region of $0\le\sigma\le1.2$ (Fig.3(a) and (b)) and $\bar{n}=1.0$ in the region of $0\le\sigma\le1.0$ (Fig.3(c) and (d)). In this figure, we consider the case that $\bar{n}=0.75$ in the region of $0\le\sigma\le1.2$ (Fig.3(a) and (b)) and $\bar{n}=1.0$ in the region of $0\le\sigma\le1.0$ (Fig.3(c) and (d)). The thick solid red lines show the minimum error probability of the atomic indirect measurement. The dotted, dashed-dotted, and dashed black lines show the minimum error probability of the feedback-excluded receiver with the photon number resolution one, two, and three, respectively. The invisibility of the feedback-excluded receiver is set to be 0.998. The solid blue lines show the Helstrom bound.}
\end{figure*}
Fig. 3 depicts the minimum error probability of the atomic indirect measurement when the message is encoded in BPSK signal. In this figure, we consider the case that $\bar{n}=0.75$ in the region of $0\le\sigma\le1.2$ (Fig.3(a) and (b)) and $\bar{n}=1.0$ in the region of $0\le\sigma\le1.0$ (Fig.3(c) and (d)). The thick solid red lines show the minimum error probability of the atomic indirect measurement. The dotted, dashed-dotted, and dashed black lines show the minimum error probability of the feedback-excluded receiver with the photon number resolution one, two, and three, respectively \cite{m.t.dimario,m.t.dimario2}. The invisibility of the feedback-excluded receiver is set to be 0.998 \cite{m.t.dimario2}. The solid blue lines show the Helstrom bound.

According to Fig. 3(a) and Fig. 3(b), the minimum error probability of the atomic indirect measurement is closer to the Helstrom bound than that of the feedback-excluded receiver with the photon number resolution one and two. When the photon number resolution is three, the minimum error probability of the atomic indirect measurement is closer to the Helstrom bound on the range of $\sigma<0.61707$. In Fig. 3(c) and Fig. 3(d), the minimum error probability of the atomic indirect measurement is closer to the Helstrom bound than that of the feedback-excluded receiver with the photon number resolution one. When the photon number resolution is two (three), the minimum error probability of the atomic indirect measurement closer to the Helstrom bound on the range of $\sigma<0.29838$ ($\sigma<0.33528$). \textit{From these results, we note that the atomic indirect measurement outperforms the feedback-excluded receiver in the aspect of the error probability when BPSK signal was sent through the phase-diffused quantum channel with not too large $\sigma$.}

\begin{figure*}
\centering
\includegraphics[scale=0.42]{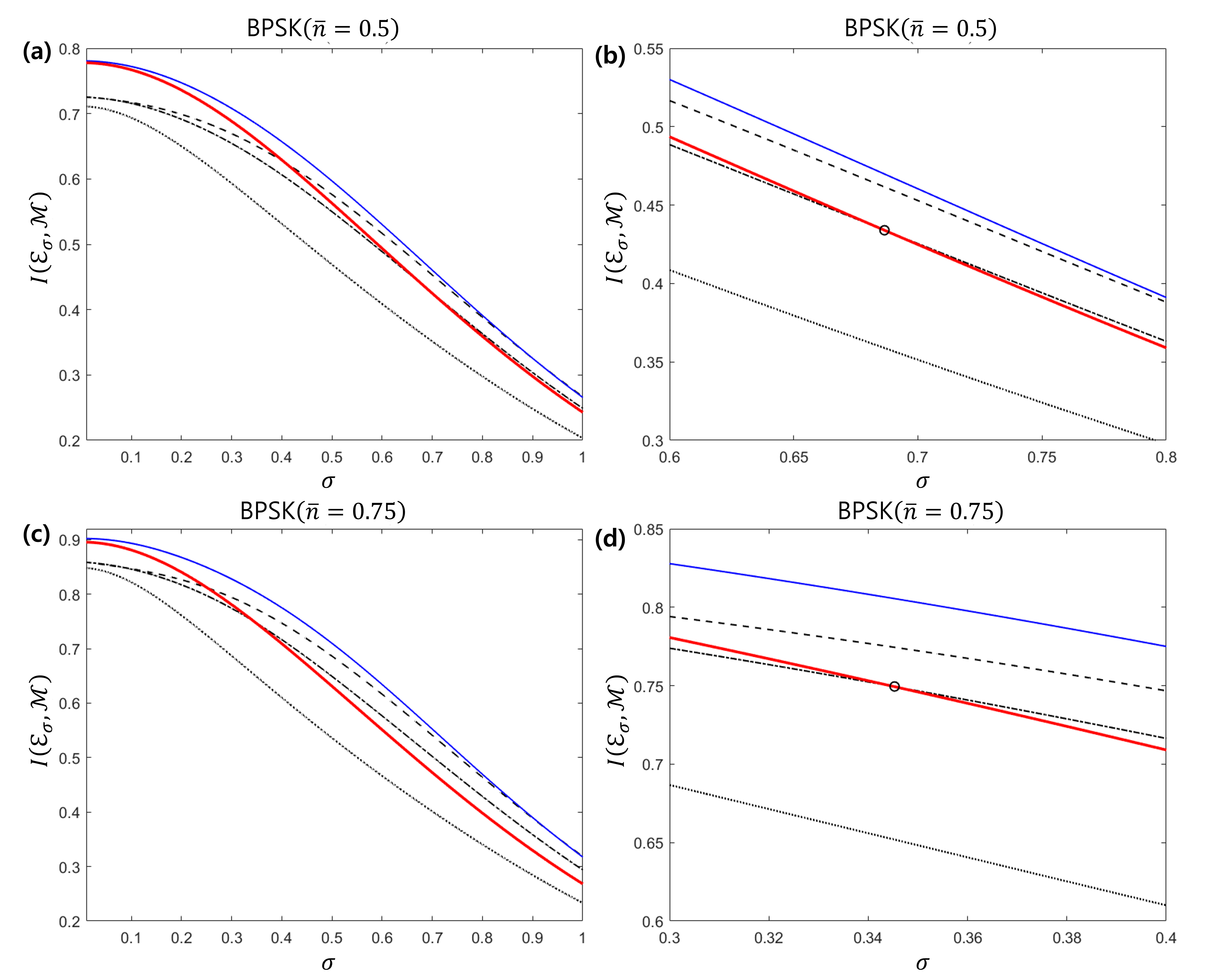}
\caption{maximal Shannon mutual information of the atomic indirect measurement when the message is encoded in BPSK signal. In this figure, we consider the case that $\bar{n}=0.5$ (Fig.4(a) and (b)) and $\bar{n}=0.75$ (Fig.4(c) and (d)) in the region of $0\le\sigma\le1.2$. The thick solid black lines show the maximal Shannon mutual information of the atomic indirect measurement. The dotted, dashed-dotted, and dashed black lines show the maximal Shannon mutual information of the feedback-excluded receiver with the photon number resolution one, two, and three, respectively . The invisibility of the feedback-excluded receiver is set to be 0.998. The solid blue lines show the accessible information.}
\end{figure*}

\subsection{Shannon mutual information}
Fig. 4 depicts the maximal Shannon mutual information of the atomic indirect measurement when the message is encoded in BPSK signal. In this figure, we consider the case that $\bar{n}=0.5$ (Fig.4(a) and (b)) and $\bar{n}=0.75$ (Fig.4(c) and (d)) in the region of $0\le\sigma\le1.2$. The thick solid black lines show the maximal Shannon mutual information of the atomic indirect measurement. The dotted, dashed-dotted, and dashed black lines show the maximal Shannon mutual information of the feedback-excluded receiver with the photon number resolution one, two, and three, respectively \cite{m.t.dimario,m.t.dimario2}. The invisibility of the feedback-excluded receiver is set to be 0.998 \cite{m.t.dimario2}. The solid blue lines show the accessible information.

According to Fig. 4(a) and Fig. 4(b), the maximal Shannon mutual information of the atomic indirect measurement is closer to the accessible information than that of the feedback-excluded receiver when the photon number resolution is one. When the photon number resolution is two (three), the maximal Shannon mutual information of the atomic indirect measurement is closer to the accessible information on the range of $\sigma<0.68653$ ($\sigma<0.4062$). In Fig. 4(c) and Fig. 4(d), the maximal Shannon mutual information of the atomic indirect measurement is closer to the accessible information than that of the feedback-excluded receiver when the photon number resolution is one. When the feedback-excluded receiver that the photon number resolution is equal to two (three), the maximal Shannon mutual information of the atomic indirect measurement is closer to the accessible information on the range of $\sigma<0.3453$ ($\sigma<0.2515$). \textit{From these results, we note that the atomic indirect measurement outperforms the feedback-excluded receiver in the aspect of the Shannon mutual information when BPSK signal was sent through the phase-diffused quantum channel with not too large $\sigma$.}

\section{Conclusion}
We have shown that the atomic indirect measurement is robust against the phase-diffusion noise.
By considering the error probability of discriminating received signal, we have shown that the atomic indirect measurement can also nearly achieve the Helstrom bound as well as the accessible information even the channel is exposed to the phase-diffused noise. Moreover, we further have shown that atomic indirect measurement outperforms the feedback-excluded receiver composed of a photon number resolving detector and maximum-a-posteriori decision rule \cite{m.t.dimario,m.t.dimario2} when the standard deviation of the phase-diffusion channel is not too large. 

Our result is more relevant to the general quantum communication protocols than the previous results that are only about the ideal quantum channels \cite{r.han,m.namkung,m.namkung2}. It is recently shown that squeezing operation can be used to reduce the phase-diffusion noise occured in the coherent state \cite{s.cialdi,g.carrara,m.n.notarnicola}. Thus, it is an interesting future work to consider the hybrid protocol composed of the atomic indirect measurement together with the squeezing operation for  robust quantum communication against the phase-diffusion noise.

\begin{acknowledgments}
We thank Dr. Donghoon Ha for his insightful discussions. This work was supported by Quantum
Computing Technology Development Program 
(NRF2020M3E4A1080088) through the National Research Foundation of Korea (NRF) grant funded by
the Korea government (Ministry of Science and ICT).

\end{acknowledgments}

\end{document}